# Plasmon-field-induced Metastable States in the Wetting Layer: Detected by the Fluorescence Decay Time of InAs/GaAs Single Quantum Dots

Hao Chen,[1,2] Junhui Huang,[1,2] Xiaowu He,[1,2] Kun Ding,[1] Haiqiao Ni,[1,2] Zhichuan Niu,[1,2,3] Desheng Jiang,[1] Xiuming Dou,[1,2,*] and Baoquan Sun [1,2,3,*]

[1] State Key Laboratory for Superlattices and Microstructures, Institute of Semiconductors, Chinese Academy of Sciences, Beijing 100083, China
[2] College of Materials Science and Optoelectronic Technology, University of Chinese Academy of Sciences, Beijing 100049, China
[3] Beijing Academy of Quantum Information Sciences, Beijing 100193, China

*To whom correspondence should be addressed: xmdou04@semi.ac.cn and bqsun@semi.ac.cn

Abstract:
We report a new way to slow down the spontaneous emission rate of excitons in the wetting layer (WL) through radiative field coupling between the exciton emissions and the dipole field of metal islands. As a result, a long-lifetime decay process is detected in the emission of InAs/GaAs single quantum dots (QDs). It is found that when the separation distance from WL layer (QD layer) to the metal islands is around 20 nm and the islands have an average size of approximately 50 nm, QD lifetime may change from approximately 1 to 160 ns. The corresponding second-order autocorrelation function $g^{(2)}(\tau)$ changes from antibunching into a bunching and antibunching characteristics due to the existence of long-lived metastable states in the WL. This phenomenon can be understood by treating the metal islands as many dipole oscillators in the dipole approximation, which may cause destructive interference between the exciton dipole field and the induced dipole field of metal islands.

Based on Fermi's golden rule, the spontaneous decay rate of dipole radiation (emitter) is directly proportional to the electromagnetic density of states $\rho(r,\omega)$ [1,2]. A modification of the emitter's electromagnetic local density of states (LDOS) will result in enhancing or inhibiting spontaneous emission as was first pointed out by Purcell [3]. The LDOS can be expressed by $\rho(r,\omega) = \sum_n f(\omega - \omega_n)|E_n(r,\omega_n)|^2$, where $E_n$ and $f$ represent the amplitude of normalized electric field and the LDOS distribution function [4]. The modification of LDOS can be realized mainly due to the coherent interference of the electromagnetic fields [2,5], for example, an emitter in front of a planar interface. If the reflected field is in phase with the emitter, LDOS at the emitter site will be high and consequently the emission will be enhanced. In contrast, if the reflected field is out of phase, the emission will be inhibited. Such kind of experimental behavior was first demonstrated by Drexhage and coworkers in the 1960s [2,6], and recently was reported in monolayer emitters [7, 8].

A metal surface with a roughness on nanometer scale can confine incident light within regions far below the diffraction limit to generate modes of localized surface plasmons (LSPs) [2,9,10], leading to a strong light-matter interaction, which has drawn intense attention in the research of physics, chemistry, materials and life sciences [11-14]. In general, light-matter interaction is realizable in the vicinity of the metal nanoparticles. It includes enhancing or quenching fluorescence radiation and a cooperative emission of light by the ensemble of emitter dipoles near metal nanoparticles [9,15-18]. Considering a system of an emitter nearby metal particles, if small particles have a radius of $R \ll \lambda$, the total electric field at the emitter site is the sum of the emitter's own field $E_0$ and dipole field of particles $E_{dip}$, $E = E_0 + E_{dip}$ [10,19]. Thus, it is proposed that in analogy with an interference effect when the emitter is in front of a planar interface, in some conditions the electric field superposition of $|E_0 + E_{dip}|^2$ may generate destructive interference (decrease of $\rho(r,\omega)$) to form metastable or dark states.

In this Letter, we report the experimental observation of abnormal long-lifetime emission in InAs/GaAs single quantum dots (QDs) through coupling between QD system and metal islands. We have measured the exciton lifetime of QD emission as a function of separation distance from QD layer to the metal islands. After the epitaxial growth, different thickness samples of separated QD films are transferred onto a substrate of the metal sheet with different surface roughness. We find that for the as grown InAs/GaAs QDs, the lifetime of QD emission is approximately 1ns, whereas for the transferred QD samples, it may increase up to approximately 160 ns. Time-resolved photoluminescence (TRPL) measurements under different excitation wavelengths and the second-order autocorrelation function measurement confirm the existence of long-lived metastable states in the wetting layer (WL). A model of classical harmonic oscillator driven by electric dipole field of metal islands is used to analyze the exciton (emitter) decay rate in the WL for a quantitative comparison with the experimental results.

The studied low density InAs/GaAs quantum dot (QD) samples were grown by molecular beam epitaxy (MBE) on a (001) semi-insulating GaAs substrate. The corresponding discrete QD emission lines can be isolated by using confocal microscopy. Details of the sample growth procedure can be found elsewhere [20]. The epitaxially grown InAs/GaAs QD samples consist of a 300 nm GaAs buffer layer, a 100 nm AlAs sacrificial layer, a thickness of D nm GaAs layer (D =10, 20, 30, 40 and 50 nm for different samples, respectively), an InAs QD layer, and a 100 nm GaAs cap layer. The QD film was separated after etching away the AlAs sacrificed layer, leaving a 100+D nm-thick GaAs film with InAs QDs. Then, each film was transferred onto a mechanically polished steel sheet (SUS301) with different surface roughness. The surface roughness of steel sheet was evaluated by 3D laser scanning confocal microscope (LEXT OLS 4000) and scanning electron microscopy (Nova Nano SEM 650).

Photoluminescence (PL) and TRPL spectra of QDs were performed at 15 K using a 640 nm pulsed semiconductor laser with a pulse length of 40 ps or a supercontinuum laser light source (YSL photonics SC-Pro) with variable wavelength and a pulse length of 100 ps as the excitation sources. A spot of about $2\ \mu m$ in diameter was focused on the sample using confocal microscope objective lens (NA: 0.55). PL spectral signal was extracted with the same microscope objective, dispersed by a $0.5\ m$ focal length monochromator and recorded by a silicon charged coupled device (CCD). TRPL spectra were measured by a time correlated single photon counting (TCSPC) device. The second-order autocorrelation function $g^{(2)}(\tau)$ measurement was carried out using the Hanbury-Brown and Twiss (HBT) setup.

Typical PL spectrum of a single QD is presented in the inset of Fig.1 (a), corresponding to the positively charged exciton ($X^+$, 908.1 nm), exciton (X, 911.3 nm) and biexciton emission (XX, 912.5 nm). These emission lines have been identified previously using the polarization-resolved PL spectra [21,22]. In addition, a broadened emission peak is observed at a wavelength of 872.2 nm and is assigned to the two-dimensional exciton emission in the WL. For the as grown QD samples, it is found that the exciton lifetime is approximately 1 ns, as evaluated by single exponential fit ($\sim \exp[-(t/\tau_1)], \tau_1 \sim 1 ns$) and shown by red line in Fig.1 (a) for $X^+$ emission. After the QD sample was transferred onto a rough steel sheet, however, its PL intensity is greatly enhanced. TRPL measurement demonstrates that except for the fast decay part of approximately 1 ns (see inset of Fig.1(b)), there is an additionally very long lifetime decay which has not been observed and reported before. TRPL curve shows a non-single-exponential decay and it can be fitted by using a stretched exponential function of $\exp[-(t/\tau_2)^\beta]$ with $\tau_2 = 160\ ns$ and $\beta = 0.8$, as plotted using red line in Fig.1(b) [23]. Analysis of TRPL curve in Fig.1 (b) shows that the decay curve consists of two independent parts. One corresponds to a fast decay process with the lifetime of $\tau_1 \sim 1\ ns$ and another is a slow one with $\tau_2 \sim 160\ ns$. In addition, $g^{(2)}(\tau)$ measurements for the as grown and transferred QD samples also demonstrate quite different result, as shown in Fig.1 (c) and (d), respectively. The former represents a typical single photon property with $g^{(2)}(0) < 0.5$, while the latter shows a bunching and antibunching characteristics, as more clearly shown in the inset of Fig.1(d) for $g^{(2)}(\tau)$ curve zoomed in at zero delay time regime. The curve can be fitted (see red line) using a three-level mode of $g^2(\tau) = a - b\exp(-|\tau|/t_1) + c\exp(-|\tau|/t_2)$ with the power-dependent fast and slow decay times of $t_1$ and $t_2$, respectively [24, 25]. The corresponding lifetimes $t_1^0 \sim 0.97$ and $t_2^0 \sim 91\ ns$ are obtained by extrapolating the data to zero excitation power (see Supplemental Material Fig.S1(a)-(d)). Thus, both results in Fig.1 (b) and (d) reveal that there should be a long-lived metastable state in the QD system [24-27]. As X and XX emissions show the characteristics similar to $X^+$ emission (see Supplemental Material Fig.S2(a)-(c)), therefore, in this report, we will just pay attention to the dynamics of $X^+$ emission.

To determine the position of metastable level in QD system, different excitation wavelengths are used to excite the QDs though excitation the InAs WL and QD excited state, respectively. The TRPL is detected at the $X^+$ emission. The results at 870 and 905 nm excitation under the saturated excitation power are presented in Fig.2 (a) and (b), respectively. They clearly show that when the light excites the WL, a similar long lifetime decay term observed in Fig.1(b) occurs, corresponding to the lifetime of approximately 145 ns. However, when the light excites the QD excited state, only a typical QD decay curve is observed, with a lifetime of approximately 1ns. Thus, experimental data clearly reveals that a long-lived metastable state is associated with the WL. In addition, if a 640 nm light excites the QD sample and the TRPL of exciton emission from WL at 872 nm is measured, the result is shown in Fig.2 (c). It shows that a short lifetime (~ 1ns) decay is followed by a long lifetime (~ 50 ns) decay. Actually, it is found that the observed lifetimes are also dependent on the excitation power (see Supplemental Material Fig.S3(a)-(d)) and metal roughness (see below). Thus, a model of three energy levels is proposed for explaining the exciton relaxation and radiation processes observed in TRPL spectra. It is schematically shown in Fig.2 (d), which describes the observed TRPL for the fast (lifetime $\tau_1$) and slow ($\tau_2$) decay curves detected at the $X^+$ emission, corresponding to the normal and metastable states in the WL, respectively.

We note that a very long lifetime in QDs can be observed after the as grown QD sample is transferred onto the rough steel sheet. Thus, the long lifetime must be related to the surface roughness of metal islands on the substrate. To examine the cause, we have fabricated different kinds of rough steel sheet as the substrate of the transferred QD films. The surface roughness of three typical polish-treated steel sheets is shown in Fig.3, evaluated by a 3D laser scanning confocal microscopy on a range of $128 \times 130 \ \mu m^2$. From this 3D result, we can get the root mean square deviation (RMS) of surface roughness of 8±4, 58±10 and 147±22 nm for the Fig.3 (a)-(c), respectively. For the studied QD samples, the thickness of GaAs layer at the bottom of QD layer (WL layer) is D = 10, 20, 30, 40 and 50 nm. After the QD films have been separated by etching and transferred onto a polished steel sheet, the distance from QDs to the steel sheet is D. In Table I, we summarize the correlation between RMS of metal sheet surface, thickness of D, and the QD lifetimes of $\tau_1$ and $\tau_2$, showing that a long lifetime can be observed only when the RMS is approximately 50 nm and D is around 20 nm. For the same RMS, the long lifetime emission vanishes when the distance D is reduced to 10 nm or increased to a distance larger than 30 nm, such as 40 or 50 nm. Furthermore, if the distance D is limited to 20 nm and QD films are transferred onto the steel sheet with a too small or too large surface roughness, e.g., 8 or 147 nm, long lifetime emission is no more found either. The TRPL spectra and fitting results are shown in Supplemental Material Fig.S4(a)-(i).

TABLE I. Experimental data for the RMS, the distance D and QD lifetimes.

| RMS (nm) | D (nm) | Lifetime (ns) |
|---|---|---|
| 8±4 | 10 | $\tau_1 = 0.95$ |
| 8±4 | 20 | $\tau_1 = 1.32$ |
| 8±4 | 40 | $\tau_1 = 1.67$ |
| 58±10 | 10 | $\tau_1 = 0.67$ |
| 58±10 | 20 | $\tau_1 = 1.1, \tau_2 = 160$ |
| 58±10 | 30 | $\tau_1 = 1.1, \tau_2 = 32$ |
| 58±10 | 40 | $\tau_1 = 2.1, \tau_2 = 9.3$ |
| 58±10 | 50 | $\tau_1 = 0.85$ |
| 147±22 | 20 | $\tau_1 = 1.39$ |

To further examine the size and roughness of surface islands on steel sheet in detail, three representative polish-treated steel sheets with RMS of 8±4 (named substrate A), 58±10 (named substrate B) and 147±22 nm (named substrate C) are analyzed using scanning electron microscopy (SEM). The results are shown in Fig. 3 (d) and (g) for substrate A, (e) and (h) for substrate B, (f) and (i) for substrate C, respectively. It is clearly shown that substrate A has a very smooth region except for several smaller islands. Substrate C has very large islands and some of them have connected to each other, forming a few hundreds of nanometer regions. For substrate B, SEM results show some discrete islands with an average size of approximately 50 nm. Actually, it is noted that only the TRPL of QD films on the substrate B (see TABLE I) shows both short and long lifetimes.

To understand the TRPL results, we model the exciton dipole (emitter) in the WL as a classical harmonic oscillator driven by the induced dipole field of metal islands [7,19]. We assume that each island serves as a hemisphere of radius $R$ and all islands are randomly distributed on the metal sheet. For small islands with $R \ll \lambda$ and in the weak-coupling regime, the polarizability $\alpha(\omega)$ of the island's response to the emitter field is modeled as a dipole oscillator characterized by [28]

$$\alpha(\omega) = \frac{\alpha_0(\omega)}{1 - \frac{ik^3 \alpha_0(\omega)}{6\pi}}, \qquad \alpha_0(\omega) = 4\pi R^3 \frac{\varepsilon - \varepsilon_m}{\varepsilon + 2\varepsilon_m} \qquad (1)$$

where $\alpha_0$ is the quasi-static polarizability, $\varepsilon$ and $\varepsilon_m$ the dielectric constant of the metal islands and surrounding medium, respectively. Assuming that the emitter dipole moment in the WL is $\vec{d}e^{-i\omega_0 t} = \hat{d}d e^{-i\omega_0 t}$, $\hat{d}$ is unit vector in the plane, which generates electric field at the metal islands is [29-31]

$$\vec{E}_0 = \frac{dk^3}{4\pi\varepsilon_0\varepsilon_m} \hat{d} \left( \frac{1}{kz} + \frac{i}{(kz)^2} - \frac{1}{(kz)^3} \right) e^{ikz} \qquad (2)$$

This field will polarize the metal islands and induce the effective dipole moment $\vec{P} = \varepsilon_0 \varepsilon_m \alpha(\omega) \vec{E}_0$ for the linear response of $\vec{P}$ to $\vec{E}_0$ [7, 19, 30]. We therefore deduce the electric field distribution of dipole moment $\vec{P}$ at the position of emitter at the WL, which can be written as [28],

$$\vec{E}_{dip} = \frac{k^3}{4\pi\varepsilon_0\varepsilon_m} \vec{P} \left( \frac{1}{kz} + \frac{i}{(kz)^2} - \frac{1}{(kz)^3} \right) e^{ikz} = \frac{dk^3 \alpha(\omega)}{16\pi^2 \varepsilon_0 \varepsilon_m} \hat{d} \left( \frac{1}{(kz)^2} + \frac{2i}{(kz)^3} - \frac{3}{(kz)^4} - \frac{2i}{(kz)^5} + \frac{1}{(kz)^6} \right) e^{i2kz}$$
(3)

When the dipole-dipole interaction between different emitters in the WL can be ignored [19], the local field at emitters is dominated by the external driving field of polarized metal islands. In this case, the total driving field $\vec{E}_{total}$ can be expressed as a superposition of free-space field $\vec{E}_0$ that would be present in the absence of metal islands and the sum of $N$ electric dipole fields of islands $\vec{E}_{dip}(j)$, $\vec{E}_{total} = \vec{E}_0 + \sum_j \vec{E}_{dip}(j)$ [10, 19]. For the substrate B, EMS is 58±10 nm. The phase difference of $N$ different dipole fields at the site of emitter can be estimated to be $k\Delta z = \frac{2\pi n \Delta z}{\lambda} \approx 0.25 \ll \pi$, with $\Delta z = 10$ nm, which satisfies the far-field approximation. Total electric dipole field is said to be an incoherent superposition of $N$ dipole fields owing to the random distribution of islands [32]. Therefore, $\vec{E}_{total}$ can be further simplified as,

$$\vec{E}_{total} = \vec{E}_0 + \sum_j \vec{E}_{dip}(j) \approx \vec{E}_0 + \vec{E}_{dip} \sum_j e^{ik\Delta z_j} \approx \vec{E}_0 + N\vec{E}_{dip} \quad (4)$$

The expression of normalized decay rate of emitters at the WL for a transition dipole in the plane can be written as [29,30],

$$\gamma/\gamma_0 = 1 + \frac{6\pi\varepsilon_0\varepsilon_m N}{dk^3} Im\{\hat{d} \cdot \vec{E}_{dip}\} =$$

$$1 + \frac{3(kR)^3 \xi N}{2} Im\left\{\left(\frac{1}{(kz)^2} + \frac{2i}{(kz)^3} - \frac{3}{(kz)^4} - \frac{2i}{(kz)^5} + \frac{1}{(kz)^6}\right) e^{i(2kz+\delta)}\right\} \quad (5)$$

where $\alpha(\omega) = 4\pi R^3 \xi e^{i\delta}$ and z is the distance from the WL (also QD layer) to the center of island. Then the modified decay rate of emitters can be evaluated using Eq. (5). Notably, in the simulation calculation, the value of the $N$ is limited to satisfy the condition of decay rate $\gamma \geq 0$ and for a random distribution of island's size, $N$ can be regarded as an average number which is not restricted to be an integer.

We show in Fig.4(a) the calculated normalized decay rate $\gamma/\gamma_0$ for $N=1$, 8 and 9.95, respectively, as a function of separation distance D (D = z-R) between the emitter in the WL and the top of the island (see inset of Fig.4(a)). The parameters used in the calculation are the wavelength of emitter in the WL of $\lambda_0 = 872\,nm$, dielectric constant of GaAs $\varepsilon_m = 13.4$ (refraction index of n=3.66), the radius $R = 25$ nm for substrate B, the dielectric constant of steel sheet $\varepsilon = -7.55 + i\,44.24$ at $\lambda_0 = 872\,nm$ as measured by Ellipsometer (Ellipsometry Solutions-M2000), $\xi = 0.93$ and $\delta = 0.95$ as calculated using dielectric constants of steel and GaAs. In Fig.4(a), we also plot five data points of the inverse of the lifetime (see TABLE I, $\gamma = 1/\tau_2$ for D=20, 30 and 40 nm, $\gamma = 1/\tau_1$ for D=10 and 50 nm) of QD emission normalized by $\gamma_0 \sim 1$ when the distance D=10, 20, 30, 40 and 50 nm, respectively. These data points are indicated by open circles. It demonstrates that the normalized decay rate $\gamma/\gamma_0$ as a function of distance D can good describe the observed experimental result of QD emission. Therefore, it is shown that the long-lived metastable states stem from the destructive interference at the emitter site (exciton in the WL) between the emitter field and dipole field of islands.

For substrates A and C, by contrast, we have not observed the long lifetime QD emission. The experimental data points (see TABLE I) are shown in Fig.4(b) by crosses for substrate A and open circle for substrate C. It is shown that the obtained lifetimes of QD emission are approximately 1ns. Similar to the estimation discussed above, we can explain why in this case there is no long-lived metastable states in the WL. It can be seen in Fig.3 (d) and (g), substrate A is a very good metal mirror and the emitters feel only reflected light. For the substrate C in Fig.3 (f) and (i), the scale of islands is comparable to the value of $\lambda_0/n(= 238\,nm)$, i.e., its area is in an order of several hundred nanometer, thus they can be approximately regarded as a reflection plane in the simulation calculation. In such an approximation, we can evaluate emitter decay rate $\gamma/\gamma_0$ when emitter is close to a metal plane. It is clearly shown by red line in Fig.4(b) that the radiation rate $\gamma$ will be suppressed at the range of a short distance when the emitter is in front of an ideal plane mirror due to the effect of destructive interference between $\vec{E}_0$ and $\vec{E}_{ref}$[30]. While for the substrates A and B, the image dipole moment is $\vec{P}' = -\left(\frac{\varepsilon-\varepsilon_m}{\varepsilon+\varepsilon_m}\right)\vec{d}$ (see inset of Fig.4(b)) [29]. This leads to that the normalized decay rate of emitters is [29,30],

$$\gamma/\gamma_0 = 1 + \frac{6\pi\varepsilon_0\varepsilon_m}{dk^3} Im\{\hat{d} \cdot \vec{E}_{ref}\} = 1 - \frac{3\xi}{2} Im\left\{\left(\frac{1}{2kz} + \frac{i}{(2kz)^2} - \frac{1}{(2kz)^3}\right) e^{i(2kz+\delta)}\right\} \quad (6)$$

where $\left(\frac{\varepsilon-\varepsilon_m}{\varepsilon+\varepsilon_m}\right) = \xi e^{i\delta}$. Figure 4(b) shows the radiation rates $\gamma/\gamma_0$ by black line for the steel reflector with $\xi = 1$ and $\delta = 0.35$. It displays that only within a short range of D the decay rate differs from the result for a perfect mirror (see red line) due to the emitters dissipating energy in the metal [1,33,34].

Actually, when the emitter decay rate $\gamma/\gamma_0 \to 0$, it can be assumed that the dark states are formed for two-dimensional exciton dipoles. In this case, based on Eq. (5), the $N$ values corresponding to $\gamma/\gamma_0 \to 0$ are approximately 9.99 (steel), 2.55 (Au) and 2.29 (Ag), as shown in Fig.4 (c). The parameters used in the calculation are $\lambda_0 = 872\ nm$ for emitter, dielectric constant of GaAs $\varepsilon_m = 13.4$, the radius $R = 25$, dielectric constant of steel $\varepsilon = -7.55 + i\,44.24$, gold $\varepsilon = -34.53 + i2.47$ and silver $\varepsilon = -35.38 + i1.94$, respectively. For a given radius $R$ (=25 nm), it can be seen in Fig4 (c) that the value of distance D is limited to be in a very narrow range for obtaining $\gamma/\gamma_0 \to 0$, such as less than 1 nm for satisfying $\gamma/\gamma_0 < 10^{-3}$. Generally, if $\gamma/\gamma_0$ takes a value of zero in Eq. (5), we can deduce a condition for observing dark states,

$$(kR)^3 N = -\frac{2}{3\xi}\left\{Im\left[\left(\frac{1}{(kz)^2} + \frac{2i}{(kz)^3} - \frac{3}{(kz)^4} - \frac{2i}{(kz)^5} + \frac{1}{(kz)^6}\right)e^{i(2kz+\delta)}\right]\right\}^{-1} \quad (7)$$

A function $(kR)^3 N \sim kz$ in Eq.(7) is plot in Fig.4 (d) by red, blue and black lines for steel, gold and silver, respectively, corresponding to the experimental condition of observed dark states. It shows that when the value of $(kR)^3 N$ approaches to the minimum, i.e., $\{kz, (kR)^3 N\} = \{1.16, 2.74\}, \{1.15, 0.69\}$ and $\{1.13, 0.62\}$ for steel, gold and silver, respectively, we can get the minimum $N$ value for observing dark states experimentally for a given radius $R$ and $\lambda$, such as the simulation results in Fig.4(c).

In conclusion, we have observed an abnormal long-lifetime emission in InAs/GaAs single QDs when the QD films are transferred onto random rough metal surfaces. The existence of a long-lived metastable states in the WL is a decisive factor for the observation of long-lifetime QD emission. The metastable states originate from destructive interference at the WL between an emitter field and induced dipole field of metal islands. In general, the emitters dissipate energy in the metal when the emitter-metal distance is close to a few nanometers. Here, long-lifetime emission in InAs/GaAs QDs is observed when the emitter-islands distance is around 20 nm, which is large enough to avoid unnecessary nonradiation losses to the metal. At the end, we designate the experimental conditions for observing dark states in the investigated system.

We acknowledge support from the National Key Research and Development Program of China (Grant No. 2016YFA0301202) and the National Natural Science Foundation of China (Grant Nos. 61827823, 61674135 and 11974342).

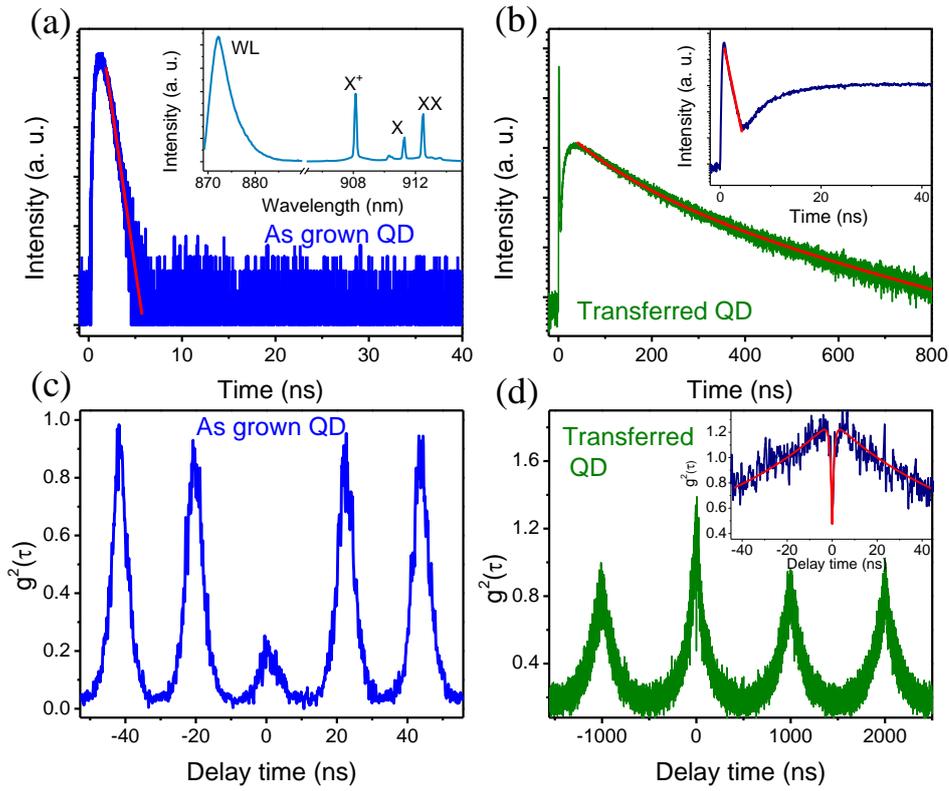

Fig.1. (a) TRPL excited by 640 nm laser and its single exponential fit (red line) for the as grown QD sample. Inset: PL spectrum of the positively charged exciton ($X^+$, 908.1 nm), exciton (X, 911.3 nm), biexciton (XX, 912.5 nm) and exciton in wetting layer (WL, 872.2 nm). (b) TRPL excited by 640 nm laser at an excitation power of 2.5 μW and its stretched exponential fit (red line) for the transferred QD sample. Inset: Zoomed in at 0-40 ns of TRPL and linear fitting to the data (red line). (c)-(d) $g^{(2)}(\tau)$ measurements for the as grown and transferred QD samples excited by 640 nm laser at an excitation power of 2 μW. Inset in (d): $g^{(2)}(\tau)$ curve zoomed in at zero delay time regime fitted using three-level model (red line).

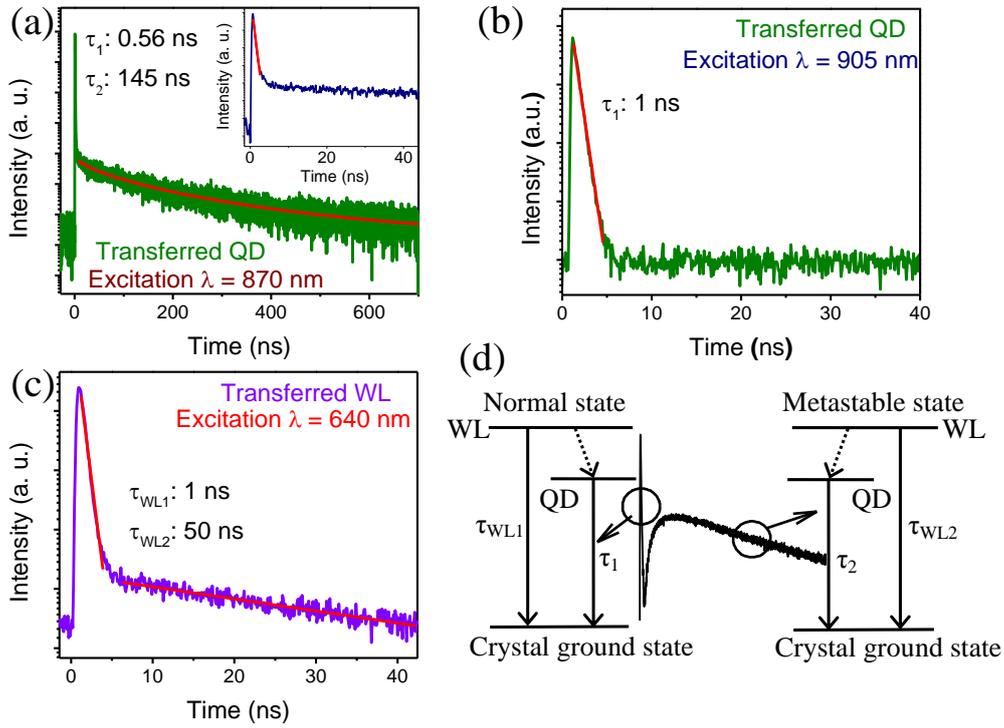

Fig.2. (a) TRPL of $X^+$ emission excited at the WL (λ=870 nm) with excitation power of 0.3 μW, yielding lifetimes of $\tau_1 = 0.56$ and $\tau_2 = 145\ ns$ for fast (see inset) and slow decay curves, respectively. (b) TRPL of $X^+$ emission excited at QD excited state (λ=905 nm) with 15 μW, yielding a lifetime $\tau_1 = 1\ ns$. (c) TRPL excited at λ=640 nm with excitation power of 0.01 μW and measured at WL emission, yielding lifetimes of $\tau_1 = 1$ and $\tau_2 = 50\ ns$, respectively. In (a)-(c), the red lines are fit to the data. (d) Schematic diagram of three-level model for illustrating fast decay (shown in left part of the figure) and slow decay (shown in right part) processes of QD emission. The TRPL curve with two decay processes is shown in middle.

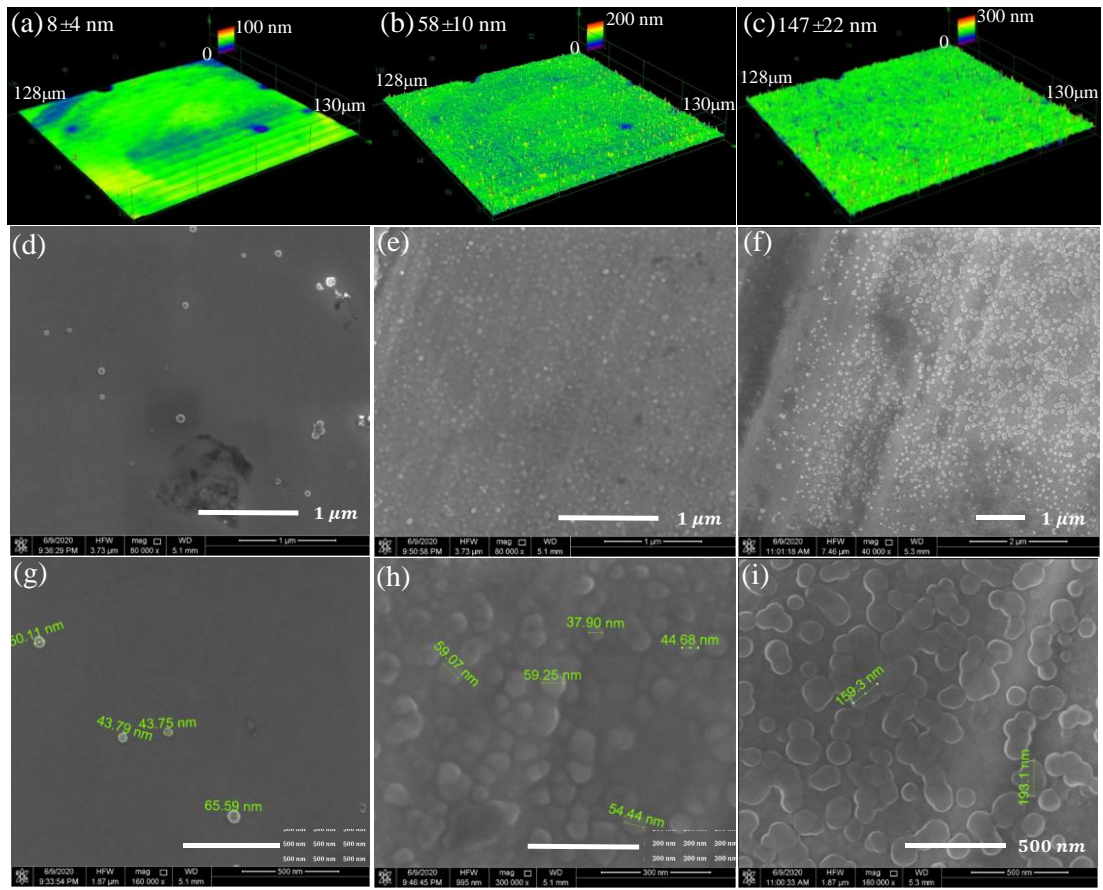

Fig.3. (a)-(c) Surface roughness of three polished steel sheets, evaluated by 3D laser scanning confocal microscopy on an area of $128 \times 130\ \mu m^2$. The RMS of surface roughness is 8±4, 58±10 and 147±22 nm, respectively. (d)-(i) SEM images for polished steel sheets with two different amplification: Substrate A, RMS = 8±4 nm (d, g), Substrate B, RMS = 58±10 nm (e, h) and Substrate C, RMS = 147±22 nm (f, i). The green numbers correspond to the size of islands and white short lines represent the scale of SEM image.

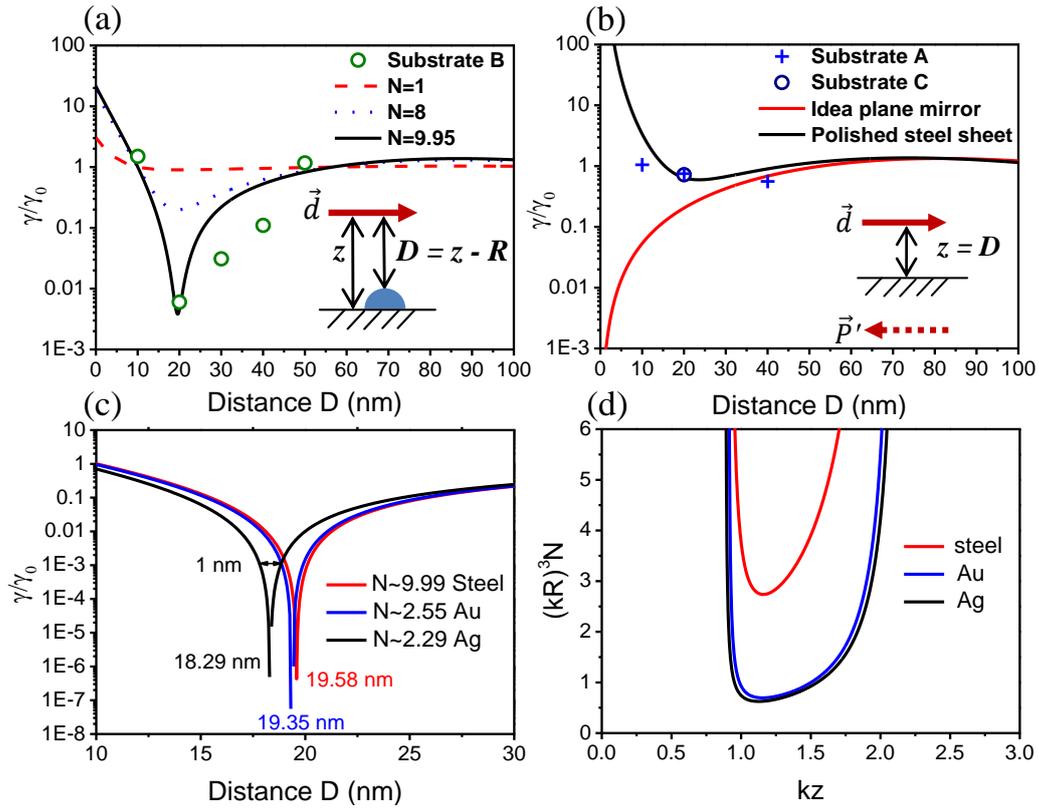

Fig.4. (a) Normalized radiation rate of emitters in the WL ($\gamma/\gamma_0$) in front of islands as a function of separation distance D (see inset of schematic diagram) based on Eq. (5) for *N*=1 (red dashed lines), 8 (blue dotted lines) and 9.95 (black line). The open circles are experimental data for substrate B. (b) Normalized radiation rate of emitters in the WL in front of idea plane mirror (red line) and polished steel sheet with dielectric constant $\varepsilon = -7.55 + i44.24$ at $\lambda_0 = 872\ nm$, as a function of distance D (see inset of schematic diagram) based on Eq. (6). The crosses and open circle represent the data for substrates A and C, respectively. (c) Normalized radiation rate of emitters as a function of D calculated using Eq.(5), showing D ~ 19.58, 19.35 and 18.29 nm for $\gamma/\gamma_0 \to 0$, corresponding to steel N~9.99 (red line), Au N~2.55 (blue line) and Ag N~ 2.29 (black line), respectively. (d) Plot $(kR)^3 N \sim kz$ of Eq.(7) for steel (red line), Au (blue line) and Ag (black line), the curves correspond to the condition of $\gamma = 0$.

Supplemental Materials：Plasmon-Field-Induced Metastable States in the Wetting Layer: Detected by the Fluorescence Decay Time of InAs/GaAs Single Quantum Dots

Hao Chen,[1,2] Junhui Huang,[1,2] Xiaowu He,[1,2] Kun Ding,[1] Haiqiao Ni,[1,2] Zhichuan Niu,[1,2,3] Desheng Jiang,[1] Xiuming Dou,[1,2,*] and Baoquan Sun [1,2,3,*]

[1] State Key Laboratory for Superlattices and Microstructures, Institute of Semiconductors, Chinese Academy of Sciences, Beijing 100083, China
[2] College of Materials Science and Optoelectronic Technology, University of Chinese Academy of Sciences, Beijing 100049, China
3  Beijing Academy of Quantum Information Sciences, Beijing 100193, China

*To whom correspondence should be addressed: xmdou04@semi.ac.cn and bqsun@semi.ac.cn

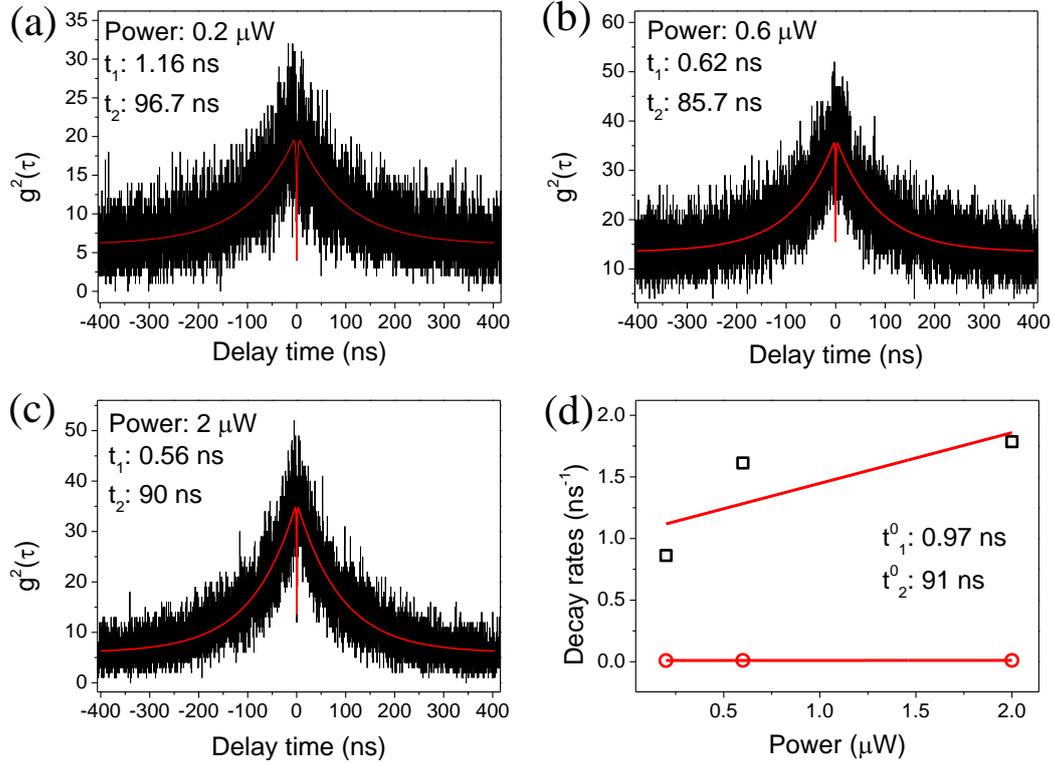

Fig.S1(a)-(c) $g^{(2)}(\tau)$ measured by using 640 nm pulsed laser excitation at the excitation power of 0.2, 0.6 and 2 $\mu W$, respectively. It shows a typical bunching and antibunching statistical characteristics. The curve can be fitted (see red line) using the three-level mode of $g^2(\tau) = a - b\exp(-|\tau|/t_1) + c\exp(-|\tau|/t_2)$ with power-dependent fast and slow decay times of $t_1$ and $t_2$ [24, 25]. The obtained $t_1, t_2$ are 1.16, 96.7 ns for (a), 0.62, 85.7 ns for (b) and 0.56, 90 ns for (c). We plot the power-dependent emission decay rate $1/t_1$ and $1/t_2$, as shown in (d). The corresponding lifetimes $t_1^0 \sim 0.97$ (corresponding to the $\tau_1$ obtained from TRPL) and $t_2^0 \sim 91\ ns$ (corresponding to the $\tau_2$ obtained from TRPL) are obtained by extrapolating the data to zero excitation power.

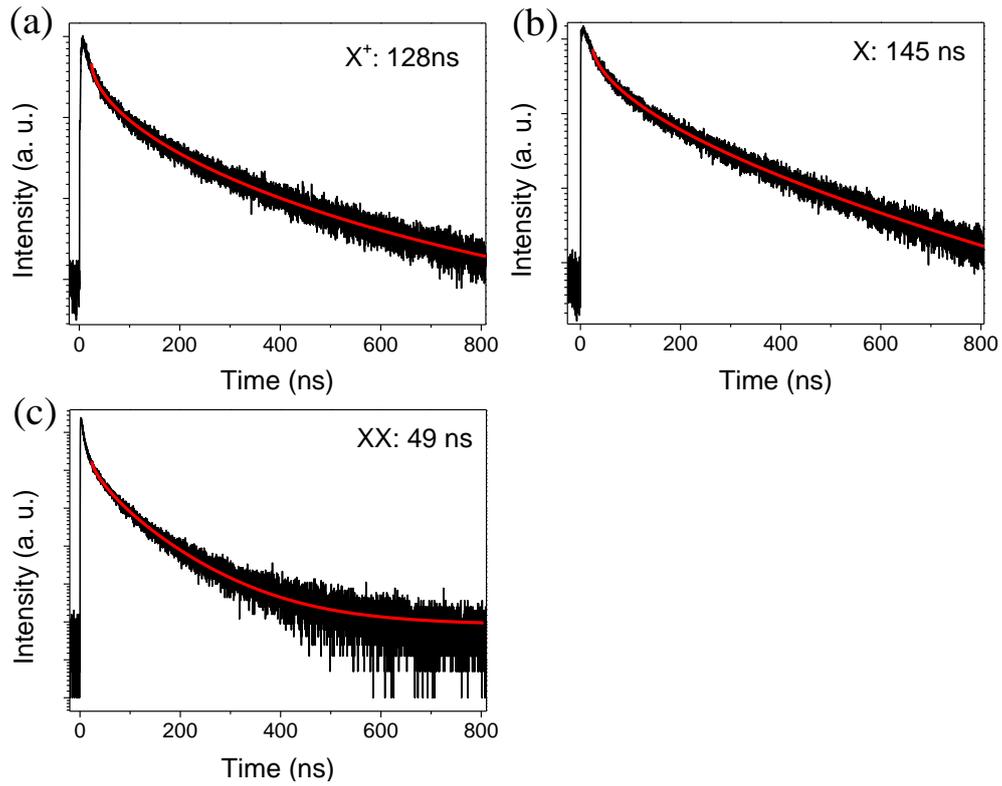

Fig.S2 (a)-(c) TRPL spectra measured at the excitation power of 0.075 μW, showing a long lifetime decay curve for $X^+$, X and XX emissions, respectively. By using a stretched exponential function fitting to the curves, the obtained decay times are 128, 145 and 49 ns for $X^+$, X and XX emissions, respectively.

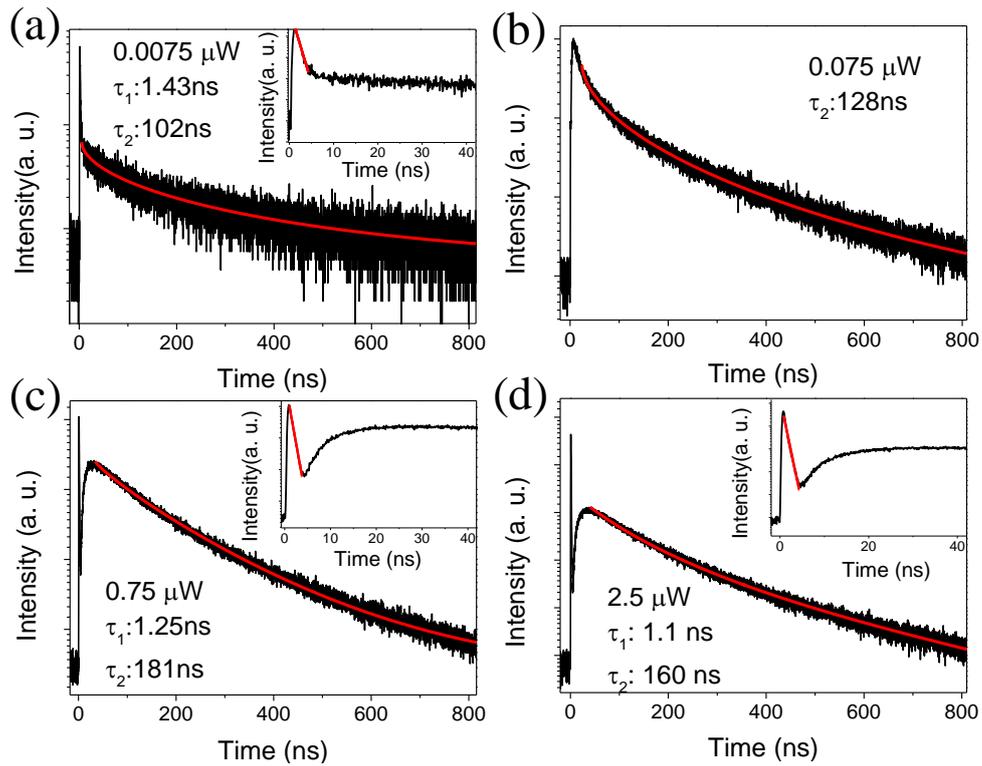

Fig.S3 (a)-(d) TRPL spectra measured at different excitation power values. The insets are shown in a much shorter time scale. It clearly shows that there are fast and slow decay times as shown in (a), (c) and (d), but in (b) the fast decay curve cannot be distinguished from the slow one. The fast decay time can be obtained by using a linear fitting to the curves in the logarithmic coordinate, they are 1.43 ns at an excitation power of 0.0075 μW as shown in (a), 1.25 ns at 0.75 μW in (c) and 1.1 ns at 2.5 μW in (d). By using a stretched exponential function fitting to the curves, the obtained slow decay times are 102, 128, 181 and 160 ns for the excitation power of 0.0075, 0.075, 0.75 and 2.5 μW, respectively. The results demonstrate the value of long lifetimes is related to the excitation power, but the detailed dependencies require further experiments.

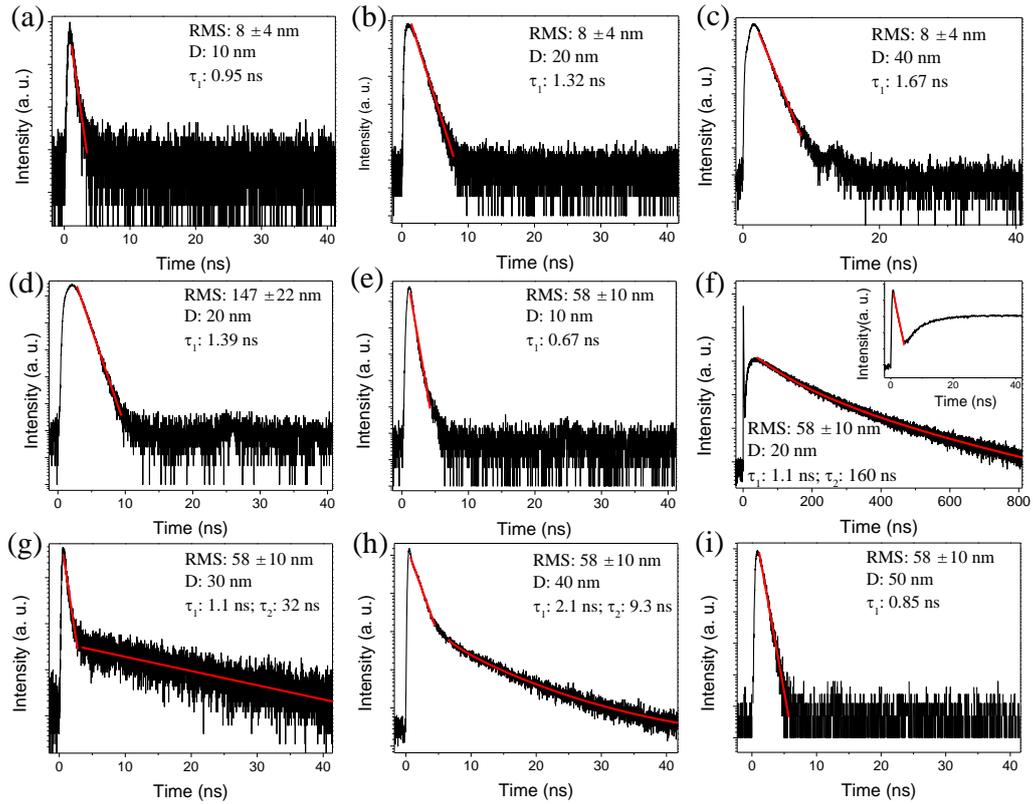

Fig.S4 (a)-(c) TRPL spectra of $X^+$ emission after the QD film was transferred onto substrate A (RMS=8±4 nm). Corresponding to the thickness D of the film is 10, 20, and 40 nm. By using linear fitting to the curves in the logarithmic coordinate, we get the lifetimes of 0.95, 1.32 and 1.67 ns for (a), (b) and (c), respectively. (d) It corresponds to the substrate C (RMS=147±22 nm), the thickness D of 20 nm and the lifetime of 1.39 ns. It clearly shows that for the substrate A and C, no long lifetime emission has been observed. The TRPL spectra of $X^+$ emission for substrate B (RMS=58±10 nm) is shown in (e, f, g, h and i). It shows that in (e) and (i) only the fast decay curves are observed and the corresponding the lifetimes are 0.67 (thickness of D =10 nm) and 0.85 ns (thickness of D=50 nm), respectively. For the thickness D of 20, 30 and 40 nm, the curves correspond to two decay times. By using a linear and a stretched exponential function fitting to the curves, the obtained fast and slow decay times are $\tau_1 = 1.1$, $\tau_2 = 160\ ns$ (f), $\tau_1 = 1.1$, $\tau_2 = 32\ ns$ (g) and $\tau_1 = 2.1$, $\tau_2 = 9.3\ ns$ (h).